\documentclass{elsart5p}

\usepackage{graphics}
\usepackage{graphicx}
\usepackage{amssymb}

\begin{document}

\begin{frontmatter}



\title{The geometrically-averaged density of states calculated from the local Green's function as a measure of localization}

\author[AA]{R.\ Wortis},
\ead{rwortis@trentu.ca}
\author[AA]{Yun Song},
\author[AA]{W.\ A.\ Atkinson}

\address[AA]{Department of Physics \& Astronomy, Trent University, 1600 West Bank Drive, Peterborough, ON, K9J 7B8, Canada}

\corauth[Wortis]{Tel: (705) 748-1011 x7714 Fax: (705) 748-1652}

\begin{abstract}
With the goal of measuring localization in disordered interacting systems, we examine the finite-size scaling of the geometrically-averaged density of states calculated from the local Green's function with finite energy resolution.  Our results show that, unlike in a simple energy binning procedure, there is no limit in which the finite energy resolution is irrelevant.
\end{abstract}


\end{frontmatter}

Significant disorder, for example due to non-stoichiometric doping, is a feature of many materials, and may in fact play a key role in many interesting behaviors, especially when combined with electronic correlations.  
The study of systems with both disorder and interactions has a long history,\cite{belitz} but it has been difficult to access the regime in which both are strong, where the physics of both Anderson localization and the Mott gap are significant.
The technique of dynamical mean field theory\cite{georges} (DMFT) has been very successful in the study of systems with strong interactions, and there has recently been interest in applying it to disordered systems.\cite{dobro,byczuk}
DMFT calculates the local Green's function, $G({\bf r},{\bf r},\omega+i\gamma)$.  How can information on localization be drawn from this?

The geometrically-averaged density of states (GADOS) appears to be ideally suited to the task.
This is the geometric average over sites of the local density of states (LDOS) which is obtained directly from the local Green's function:
\begin{eqnarray}
\rho_g(\omega) &=& \left[ \prod_{i=1}^N \rho({\bf r}_i,\omega) \right]^{1/N},
\\
\rho ({\bf r}_i,\omega) &=& -{1 \over \pi} {\rm Im} G({\bf r}_i,{\bf r}_i, \omega+i\gamma)
\end{eqnarray}
which is exact in the limit $\gamma \rightarrow 0^+$.
In an infinite system, $\rho_g(\omega)$ at the Fermi level is an order parameter for the Anderson transition, being nonzero when states at the Fermi level are extended and zero when they become localized.  
This is because when wavefunctions at a given energy become localized in real space, the LDOS at that energy becomes discrete: localized wavefunctions at the same energy are well separated in position so the LDOS at that energy goes to zero far from the wavefunction centers.

To consider specific disorder configurations, it is necessary to work with finite size systems, and in this case finite-size scaling analysis of the GADOS must be performed.  
This procedure is well studied in non-interacting systems for which single-particle wavefunctions can be calculated, providing perfect energy resolution.
In a numerical calculation on an interacting system, however, the Green's function is obtained for finite $\gamma$.
We have explored the impact this has on the scaling of the GADOS in the context of a noninteracting model 
\begin{eqnarray}
{\cal H} &=& \sum_i |i\rangle \epsilon_i \langle i| - t \sum_{\langle i,j \rangle} |i\rangle \langle j|,
\end{eqnarray}
with site energies, $\epsilon_i$, distributed with probability
$P(\epsilon_i) = \Theta (W/2 - |\epsilon_i|) /W$.
The width of the potential distribution, $W$, parameterizes the strength of the disorder.
The model is solved exactly by numerical diagonalization using the LAPACK libraries.
In earlier work,\cite{saw} 
we calculated the local density of states within a finite energy resolution $\gamma$ using a simple binning procedure:
\begin{eqnarray}
\rho({\bf r}_i, \omega) &=& \sum_{\alpha} |\Psi_{\alpha}({\bf r}_i)|^2 
{\Theta (\gamma/2 - |\omega-E_{\alpha}|) \over \gamma}.
\label{box}
\end{eqnarray}
Each state which falls in the energy window is assigned equal weight.
In this case, we demonstrated that the GADOS scales as a function of an effective length scale:
$L_{eff} = 1/(1/L + 1/\ell_{\gamma})$ where
$\ell_{\gamma}$ acts as an effective inelastic mean free path, limiting the relevant system size.  $\ell_{\gamma}$ can also be seen as the system size divided by the average number of states in a bin: $\bar{n}=\gamma/\Delta$ where in one dimension $\Delta=1/(\rho L)$ is the average energy spacing between states and $\rho(\omega)$ is the density of states.  

Here we model more accurately the way the Green's function introduces finite energy resolution:
\begin{eqnarray}
\rho({\bf r}_i, \omega) &=& \sum_{\alpha} |\Psi_{\alpha}({\bf r}_i)|^2 
{\gamma \over \pi [ (\omega-E_{\alpha})^2 + \gamma^2]}.
\label{lor}
\end{eqnarray}
All results shown are at the band center on one-dimensional systems with disorder strength $W=1$, for which all states are localized.  

\begin{figure}[!ht]
\begin{center}
\includegraphics[width=0.45\textwidth]{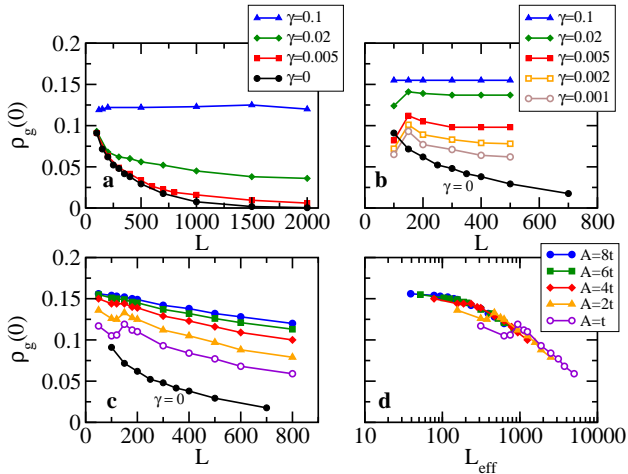}
\end{center}
\caption{The geometrically-averaged density of states at the band center in a one-dimensional system with disorder strength $W=1$.  All points represent averages over at least 1000 disorder configurations.  $L$ is the system size, and $L_{eff}=L/(\rho A)$ where $\rho$ is the density of states.  
(a) Square energy bin (Eqn.\ (\ref{box})) with constant width $\gamma$.
(b) Lorentzian energy bin (Eqn.\ (\ref{lor})) with constant $\gamma$.
(c) and (d) Lorentzian energy bin with $\gamma=A/L$.} \label{fig1}
\end{figure}

Fig.\ 1(a) shows the GADOS as a function of system size for three non-zero values of $\gamma$ using the square binning in Eqn.\ (\ref{box}) reported earlier.\cite{saw} 
Fig.\ 1(b) shows the GADOS as a function of system size for the same $\gamma$ values using the Lorentzian binning in Eqn.\ (\ref{lor}).  Two features are immediately apparent:  the lack of exponential decay and the overall inflation of the GADOS.
With the Lorentzian bin, {\it all} states contribute to the sum in Eqn.\ (\ref{lor}) with rapidly decreasing weight away from $\omega$.  
Despite the small weight assigned to them, the many states outside the peak of the Lorentzian have a dramatic impact on the GADOS.  This is because when these states are centered, in real space, at positions with otherwise small LDOS they can significantly increase the LDOS.  Even a modest increase in the {\it minimum} values of the LDOS causes a dramatic increase in the GADOS.

A second important distinction between the two binning schemes arises in the limit $\gamma < \Delta$.  
In this limit, for each $\omega$ there will be disorder configurations for which (i) in the square bin case there are no states in a bin and (ii) for the Lorentzian bin case no state has significant weight.  
In the case of the square bin, we removed these bins from the average.  
However, there is no systematic way to do this in the Lorentzian case.  
The inclusion of these states dramatically lowers the configuration averaged LDOS and hence the GADOS even below the $\gamma=0$ value.

It appears to be possible to account for the behavior of $\rho_g$ with Lorentzian binning within the framework of an effective length; however, a different form of $L_{eff}$ is required.
In the square bin case, in the limit of small $\gamma$, $L_{eff}=L$, because by construction there is always at least one state in each bin.
In the Lorentzian bin case, the average number of states in a bin may be less than one.
$L_{eff}$ is now simply equal to $L/\bar{n} = L \Delta/\gamma = 1/(\rho \gamma)$.
This is consistent with the nearly size independent results when a fixed value of gamma is used.
Furthermore, 
Fig.\ 1(c) shows the GADOS as a function of system size using bin widths $\gamma = A/L$.
Fig.\ 1(d) presents the same data plotted versus 
\begin{eqnarray}
L_{eff}&=& {L \over \rho A},
\end{eqnarray}
showing scaling.  
The key point is that there is no limit in which the finite energy resolution is irrelevant.

In conclusion, there appears to be no finite-size scaling of the GADOS calculated from the local Green's function with fixed energy resolution, and the scaling which occurs when $\gamma \propto 1/L$ arises at least in part from physics unrelated to localization.
We suggest that other measures of localization, such as the inverse participation ratio, which do not share the extreme sensitivity to small values of the LDOS may be more effective than the GADOS in interacting systems, although changes to the scaling due to finite energy resolution will still need to be addressed.

\section{Acknowledgement}
This work is supported by NSERC and Trent University.

\end{document}